\documentstyle[11pt,moriond,epsfig]{article}

\def\Journal#1#2#3#4{{#1} {\bf #2}, #3 (#4)}

\def\PRL{\em Phys. Rev. Lett.}
\def\PRD{{\em Phys. Rev.} D}

\def\mco{\multicolumn}

\def\be{\begin{equation}}
\def\ee{\end{equation}}
\def\bea{\begin{eqnarray}}
\def\eea{\end{eqnarray}}

\def\D0{D\O}
\def\Gev2{GeV/c$^2$}
\newcommand{\met}     {\mbox{$\not\!\!{E_T}$}}
\newcommand{\etal}{{\sl et al. }}
\begin{document}
\begin{flushright}
{FERMILAB-CONF-01/119-E}
\end{flushright}
\vspace*{4cm}
\title{HIGGS SEARCHES AT THE TEVATRON}

\author{L. MONETA \\
(for the CDF and \D0 Collaborations)}

\address{University of Geneva, DPNC\\
CH-1211 Geneva, Switzerland }

\maketitle\abstracts{
This paper describes the searches for the Higgs boson performed by the CDF and 
\D0 collaborations at the Tevatron $p \bar p$ Collider using the data
collected in the 1992-95 run. 
Searches for standard model Higgs and as well for neutral and 
charged minimal SUSY Higgs bosons are also presented.    
No signal has been observed and limits are set for 
production cross sections.   
}

\section{Introduction}

The Standard Model (SM) predicts the existence of the Higgs boson which is
responsible of the electroweak symmetry breaking. 
The mass is not predicted by the theory but 
precision electroweak measurements 
constraint the Higgs mass below 212 \Gev2~\cite{ew_fit}(at 95\% CL). Direct
searches at LEP~\cite{higgs_lep} exclude $< 113.5$ \Gev2. In  
addition,  
an excess of events consistent with a Higgs boson with mass of
$115$ \Gev2 is observed at LEP~\cite{higgs_lep}, 
but it is not statistically conclusive. 
\par                                  
In the minimal supersymmetric extension of the SM (MSSM), the 
electroweak symmetry
breaking of the Higgs sector requires the presence of 
three neutral ($h$, $H$ and $A$) and two charged ($H^{\pm}$) scalar Higgs bosons. The mass of the lightest Higgs, $h$, is constrained by the model to 
$M_h \le 130$ \Gev2, while the other masses are expected to be larger.     

\section{Searches for the Standard Model Higgs} 

The dominant Higgs production processes at Tevatron are the gluon fusion, 
$gg \to H$, and associated production with a $W$ or $Z$ boson, $q \bar q \to
W/ZH$. 
The gluon fusion has higher cross sections but it is very difficult to
distinguish from the overwhelming QCD background in the 
light mass region ($m_H \le 130$ \Gev2), when $H \to b
\bar b$ is the dominant decay channel. In the case of associated production,  
the vector boson provides a handle to 
suppress the background, but 
the cross section, $\sigma( p \bar p \to W/Z H ) \approx  300$ fb for $m_H = 110$
\Gev2, is approximately
20 times smaller than the $t \bar t$ cross section. 
According to the decay mode of the vector bosons, four different signatures 
are considered: $\ell \nu b \bar b$, $\nu \nu b \bar b$ , $\ell^+ \ell^- b \bar b$ 
and  $jj b \bar b$ ($\ell = e, \mu$). 
\par
The search strategy is similar for all decay channels. After identifying 
the $W$ or $Z$ through its decay, at least two extra jets from the  
$H \to b \bar b$ decay are required and, depending on the analysis,    
one or both jets, must be tagged as b jets.     
Two b-tagging techniques, have been developed and applied successfully by CDF
and \D0 in the discovery of the top quark. 
The first methods look for leptons within a jet 
(soft lepton tagging) exploiting the semileptonic decays of the b hadrons. 
The second method (secondary vertex or jet probability) is more powerful and  
is based on the presence of the displaced decay vertex 
within the jet, arising from the relative long B lifetime. 
Only CDF, equipped in the past run with a silicon vertex detector, could 
use this second technique. 
The Higgs mass is reconstructed using the two b tagged jets or the two most
energetic jets. Jet energy resolution plays as well a major role to
discriminate signal against background. 
\par
Both CDF~\cite{cdf_lnu} and \D0~\cite{d0_lnu} search for the 
SM Higgs in the $WH \to \ell \nu b
\bar b$ channel.
These events are reconstructed 
by requiring one isolated electron and muon with $p_T > 20$ GeV/c,   
transverse missing energy $\met > 20$ GeV and 
two jets with $E_T > 15$ GeV and $\eta < 2$. 
CDF distinguishes when one or both jets are identified as b jets using the secondary vertex and the soft lepton tag, while \D0 requires at least one soft muon b tagged jet.     
Major remaining sources of background are $W$ produced in association
with heavy flavours ($W b \bar b$, $W c \bar c$), fake b tags, and top
events.    
%
%
\par
A preliminary search in the $ZH \to \nu \nu b \bar b$ channel is performed 
by CDF. 
The signature consists in the presence of
 large missing transverse energy  
($\met > 40$ GeV), no charged lepton and two or three jets. 
QCD background from mismeasured $\met$  is reduced  
by requiring $\Delta \Phi_{min}(\met, jet) > 60^0 $ and background
 from QCD dijet events is reduced with a cut on  
 $\Delta \Phi (jet_1, jet_2) < 120^0 $. 
Independent analyses with single and double b-tag are
 performed. After b-tagging, the remaining backgrounds are QCD, $W b \bar b$,
 $W c \bar c$ top and dibosons productions. 
%
%
%
\par
The $ZH \to \ell^+ \ell^- b \bar b$ channel is the cleanest given the presence of two leptons with  
the possibility of applying the  $Z$ mass constraint but it is limited by the 
low $BR(Z \to \ell^+ \ell^-)$.  
CDF has performed a preliminary search looking for two high $p_T$ electrons or muons 
($p_T > 20,10$ GeV/c) with invariant mass, $76 < M_{\ell \ell} < 106$ \Gev2. 
Two or three jets are required, and at least one must be b-tagged. 
The remaining backgrounds are 
$Z b \bar b$, $Z c \bar c$ and fake tags. 
%
\par
The $Z/WH \to q \bar q b \bar b$ channel contributes to $\sim  50$ \% of all decays, but suffers from the
high QCD multi jet production. 
The CDF search~\cite{cdf_qq} starts by accepting events with 4 or more jets, among which at
least two are tagged as b-jets by the secondary vertex algorithm.  
In addition, the $p_T$ of the two b-jets system is required to be 
$p_T(b \bar b) > 50$ GeV/c. 
The background is dominantly QCD, but contributions from
fake b-tags, top and $Z +$ jets are also present. 
%
%
\par
\begin{figure}
\mbox{\epsfig{file=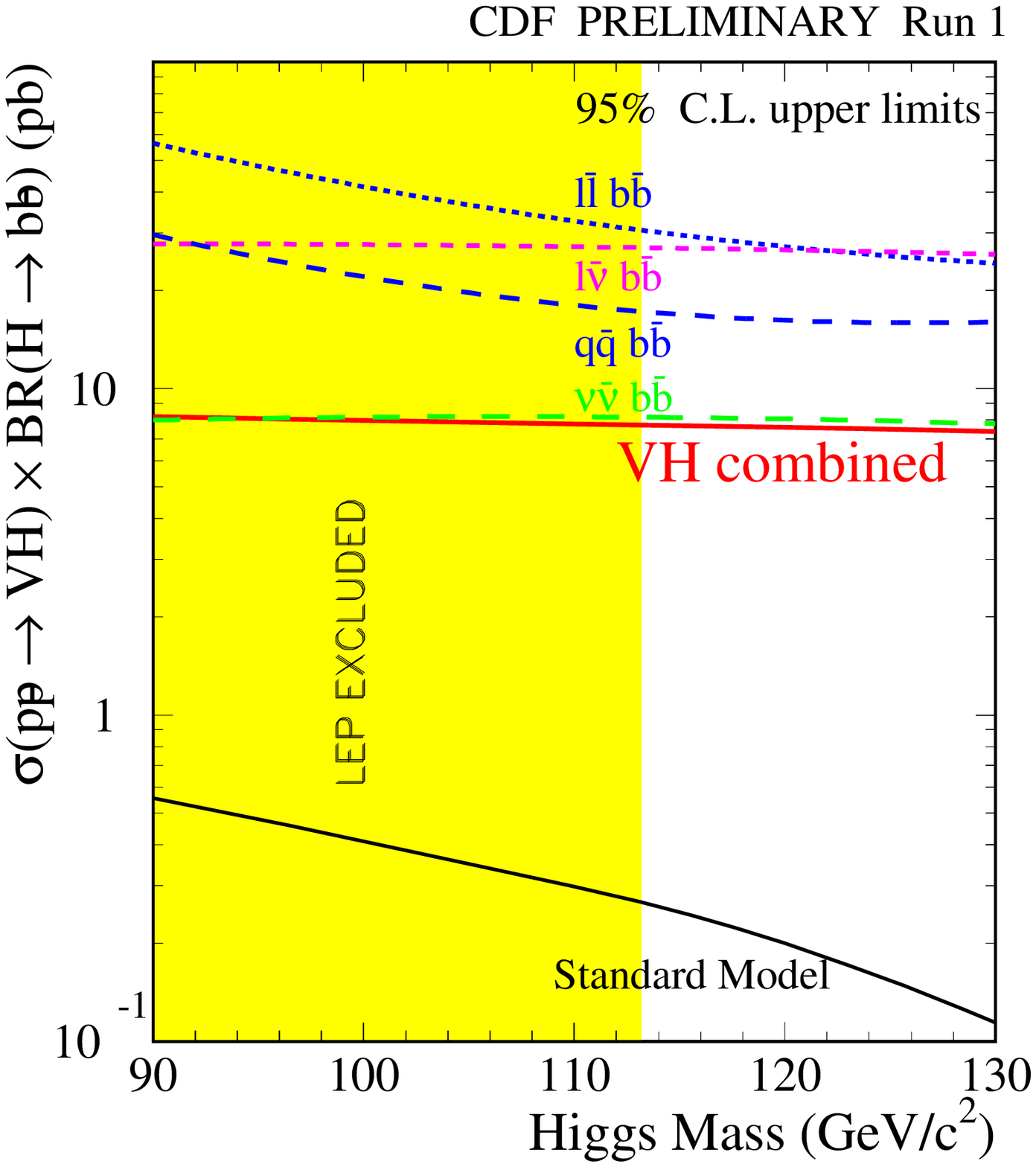, width=0.4\textwidth} \hspace{0.02\textwidth}
\epsfig{file=h_d0_limit.eps, width=0.5\textwidth}}

\caption{95\% CL upper limits on production cross section times branching ratios as function of Higgs mass from CDF (left) and \D0 (right). The combined 
CDF limit is calculated by fitting the product of the individual likelihood for each channel, taking into account correlation between systematic uncertainties.
\label{smh_plot}}
\end{figure}

The results of the CDF searches are summarised in
table~\ref{smh_results}, where the estimated acceptance for each
channel and the number of events observed in the data are reported 
together with the total background predictions after 
the $W/Z + b \bar b$ selection. The observed events are consistent with the expectations and 95\% CL limits on the production cross section times branching ratio are derived from a maximum likelihood fit to the dijet mass distributions. Figure~\ref{smh_plot} shows the results for each channel and their combination.

\begin{table}[t]
\caption{Table summarising the results for the SM Higgs searches showing for
  each channel the corresponding luminosity, branching ratio $\times$
  acceptance, expected background and observed events.
\label{smh_results}}
\begin{center}
\begin{tabular}{|cc|c|c|ccc|c|c|}
\hline
\mco{2}{|c|}{Channel} & Experi- & $\int \cal L$ & 
\mco{3}{|c|}{BR $\times$ Accept.(\%) for $M_H$ (\Gev2)} 
& Expected & Observed \\
 & b-tags & ment & (pb$^{-1}$) &  90 & 110 & 130  & background & events \\ \hline 
$\ell \nu b \bar b$ & 1 & CDF\cite{cdf_lnu} & 106 & 
$.55 \pm .14$ & $.74 \pm .18$ & $.89 \pm .22$ & $30 \pm 5$ & 36    \\

$\ell \nu b \bar b$ & 2 & CDF\cite{cdf_lnu} & 106 & 
$.23 \pm .06$ & $.29 \pm .07$ & $.34 \pm .09$ & $3.0 \pm 0.6$ & 6    \\

$\ell \nu b \bar b$ & 1 & \D0\cite{d0_lnu} & 106 & 
$.30 \pm .02$ & $.36 \pm .02$ & $.44 \pm .03$  & $25.5 \pm 3.3$ & 27    \\

$\nu \nu b \bar b$ & 1 & CDF  & 88 & 
$.59 \pm .12$ & $.69 \pm .14$ & $.86 \pm .17$ & $39.2 \pm 4.4$ & 40    \\

$\nu \nu b \bar b$ & 2 & CDF  & 88 & 
$.37 \pm .08$ & $.44 \pm .11$ & $.53 \pm .11$ & $3.9 \pm 0.6$ & 4    \\

$\ell \ell b \bar b$ & 1 & CDF & 106 & 
$.14 \pm .03$ & $.20 \pm .04$ & $.19 \pm .04$ & $3.2  \pm 0.7$ & 5     \\

$q \bar q b \bar b$ & 2 & CDF\cite{cdf_qq} & 91 & 
$1.3 \pm .4$ & $2.2 \pm .6$ & $3.1 \pm .8$ & $594 \pm 30$ & 589   \\

\hline
\end{tabular}
\end{center}
\end{table}

\section{Search for  MSSM Neutral Higgs} 

\begin{figure}[t]
\mbox{
\epsfig{file=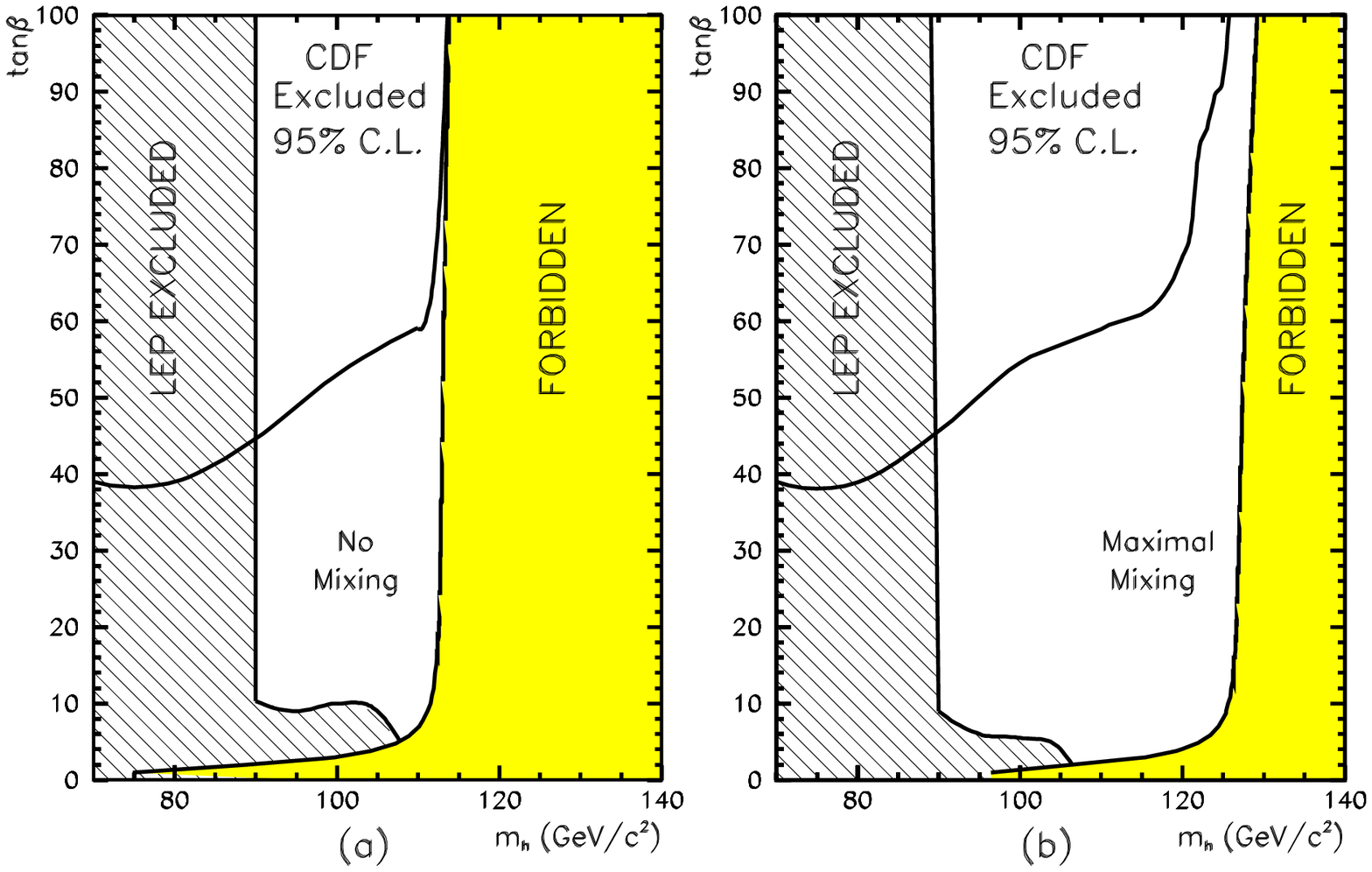, width=0.5\textwidth} 
\epsfig{file=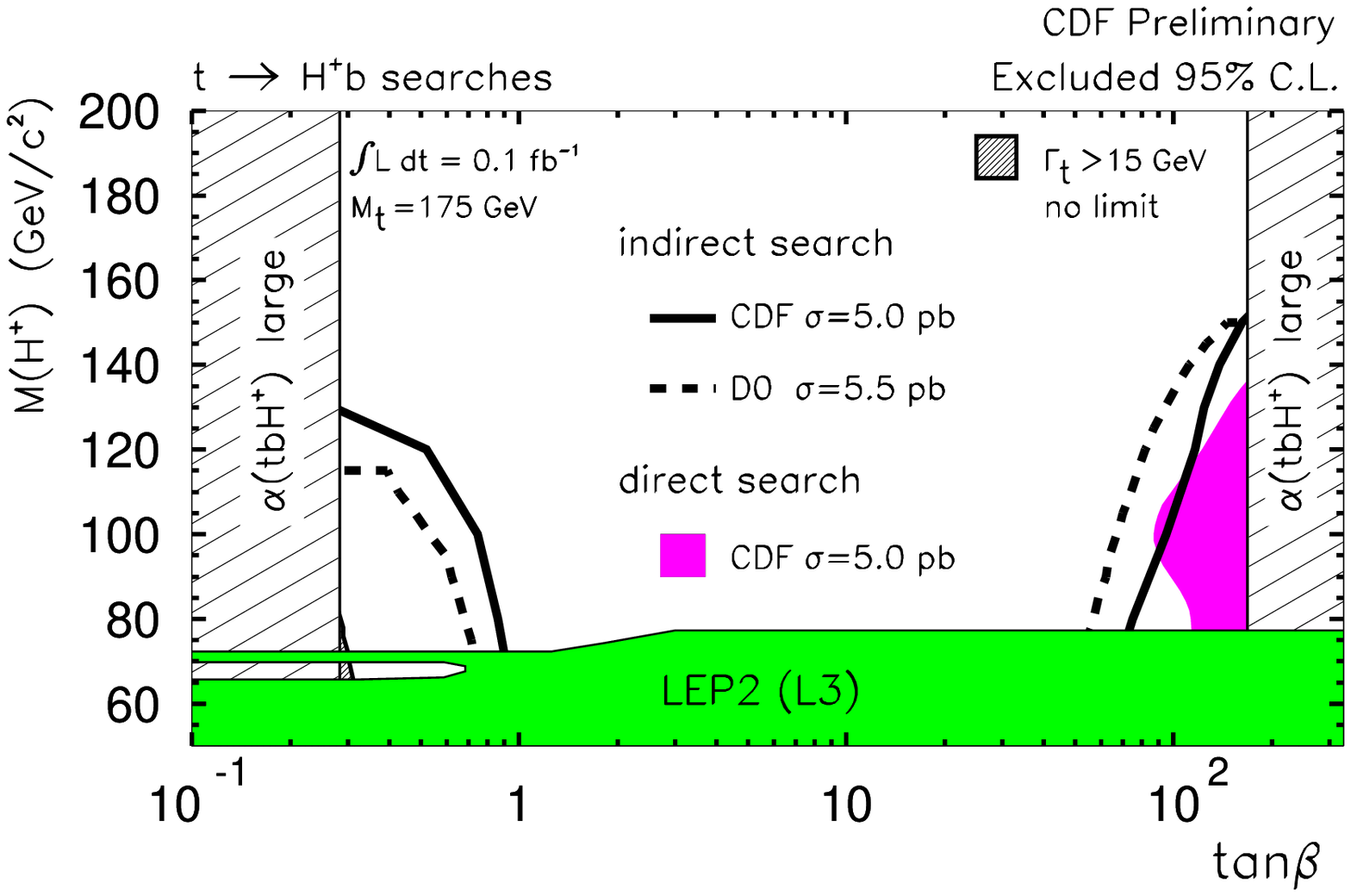, width=0.5\textwidth}
}
\caption{95\% CL exclusion region in the $\tan \beta$ vs $M_h$ plane from the
  CDF MSSM neutral Higgs search (left) and  exclusion region in $\tan \beta$ 
vs $M_{H^{\pm}}$ plane for the charged Higgs searches(right) from CDF and 
\D0. 
\label{h_mssm}}
\end{figure}

CDF has searched for a neutral MSSM Higgs ($\phi$) produced in association with $b \bar b$, process $p \bar p \to b \bar b \phi \to  b \bar b b \bar b$,
using 91 pb$^{-1}$ of data~\cite{cdf_bbh}. 
This process can be particularly important in the MSSM parameter region when the Yukawa couplings between the Higgs scalars and the $b$ quarks are enhanced with respect to 
the SM. Events are selected by requiring four jets with at least three of them
b tagged and imposing Higgs mass dependent
$E_T$ cut. In addition, a cut on
the b-tagged dijet angular distribution reduces the QCD background with $g \to
b \bar b$. The observed number of events is in agreement with SM contributions,
consisting mainly of QCD heavy flavours, mistags, top and $Zjj$ events. Excluded regions are set in the $\tan \beta$ vs $M_h$ and $m_A$ plane, for
two stop mixing scenario (no mixing and maximal mixing) and for a SUSY mass
scale of 1 TeV. As shown in figure~\ref{h_mssm}, the CDF limits probe a region
of parameter space, for $\tan \beta$ larger than 35, which could not be accessible at LEP.

\section{Search for charged Higgs from the decay of the top quark}
\label{subsec:h+}

Searches for charged Higgs are performed by CDF and \D0 looking for top quark decay to charged Higgs ($t \to H^+ b$), when $m(H^+) < m_t - m_b$. This search is sensitive only for low and large  $\tan \beta$ values, when the $BR(t \to H^+ b)$ dominates. 
The charged Higgs decays almost 100\% in $\tau \nu$ for $\tan \beta > 1$ and hadronically in $cs$ for low  $\tan \beta$ values. 
Direct searches 
from CDF~\cite{direct_h+}
look  for $H^+ \to \tau \nu$ using $\tau$ identification via its 
hadronic decays and
they are thus sensitive only in the large $\tan \beta$ region.  
Indirect searches, performed by CDF and \D0~\cite{indirect_h+}, are more 
powerful and look for suppression of SM $t \bar t$ decays, caused by $t \to
H^+ b$ decays. 
Observed rates of dilepton and lepton + jets events in the top sample depends 
on $BR(t \to H^+ b)$, while the  SM predictions for $\sigma_{t \bar t}$ 
are independent $H^{\pm}$ production. 
Exclusion regions in the $M(H^+)$ vs $\tan \beta$ plane have been
then derived with sensitivity in both  low and high   $\tan \beta$ 
region as shown in figure~\ref{h_mssm}.    

\section{Conclusion}

Results on Higgs searches from the data collected at Fermilab during 1992-95
have been reported. No evidence of signal has been found and 95\% CL limit
have been derived.  
Searches for the SM Higgs boson are presently limited 
by statistics at approximately $< 8$ pb, 30 times higher than the SM values. 
These searches are however the starting
points for the analyses of the next run. 
The increased integrated luminosity and the detector upgrades will allow
the possibility of discovering the Higgs bosons in the intermediate mass
range~\cite{run2}.
Results on searches for MSSM neutral Higgs, via $p \bar p \to b
\bar b \phi \to b \bar b b \bar b$, and charged Higgs produced from top quark
decays, are also presented. 
Limits are derived as function of the MSSM parameters, 
probing region inaccessible at LEP.

\section*{Acknowledgments}
I would like to thank all members of the CDF Exotic Physics group and the \D0 
New Phenomena group for providing me the different material shown in this report.


\section*{References}

\begin{thebibliography}{99}


\bibitem{ew_fit} T. Kawamoto, these proceedings.

\bibitem{higgs_lep} A. Okpara, these proceedings. 

\bibitem{cdf_lnu} F. Abe \etal (CDF Collaboration), \Journal{\PRL}{79}{3819}{1997}.

\bibitem{d0_lnu} S. Abachi \etal (\D0 Collaboration), Fermilab-Conf-96/258-E



\bibitem{cdf_qq} F. Abe \etal (CDF Collaboration),  \Journal{\PRL}{81}{5748}{1998}.

\bibitem{cdf_bbh} T. Affolder \etal (CDF Collaboration), \Journal{\PRL}{86}{4472}{2001}.

\bibitem{direct_h+} T. Affolder \etal (CDF Collaboration), \Journal{\PRD}{62}{12004}{2000}.

\bibitem{indirect_h+} B. Abbott \etal (\D0 Collaboration),  \Journal{\PRL}{82}{4975}{1999}.
\bibitem{run2} A. Lucotte, these proceedings.

\end{thebibliography}
\end{document}